\documentclass[showpacs,preprintnumbers,amsmath,amssymb,twocolumn,superscriptaddress,prb]{revtex4}
\usepackage{graphicx}
\usepackage{mathptm}
\usepackage{amsfonts}
\usepackage{amsmath, amsthm, amssymb}
\usepackage{dsfont}
\usepackage{times}
\usepackage{subfigure}

\newcommand{\e}[1]{\mathrm{e}^{#1}}
\newcommand{\bq}{\begin{equation}}
\newcommand{\eq}{\end{equation}}

\newcommand{\vi}{\mathbf{i}}
\newcommand{\vj}{\mathbf{j}}

\newcommand{\eg}{\textit{e.g. }}%[syn: f.eks., for example, for instance]
\newcommand{\etal}{\emph{et al.}}
\def\i{\mathrm{i}}

\begin{document}
\title[Spin-active interfaces and unconventional pairing in half-metal$\mid$superconductor junctions]
{Spin-active interfaces and unconventional pairing in half-metal$\mid$superconductor junctions}
\author{Jacob Linder}
\affiliation{Department of Physics, Norwegian University of
Science and Technology, N-7491 Trondheim, Norway}
\author{Mario Cuoco}
\affiliation{CNR-SPIN, I-84084 Fisciano (Salerno), Italy}
\affiliation{Dipartimento di Fisica "E.R. Caianiello",
Universit\'a di Salerno, I-84084 Fisciano (Salerno),
Italy}
\author{Asle Sudb{\o}} \affiliation{Department of Physics,
Norwegian University of Science and Technology, N-7491 Trondheim,
Norway}

\date{Received \today}
\begin{abstract}
\noindent We study the physical properties of a half-metallic
ferromagnet$\mid$superconductor (HM$\mid$S) bilayer, allowing for
an arbitrary bulk pairing symmetry of the superconductor and
spin-dependent processes at the interface. In particular, we study
how the possibility of unconventional pairing such as $p$- and
$d$-wave and a spin-active interface influence the \textit{(i)}
conductance spectra, \textit{(ii)} proximity effect, and
\textit{(iii)} local density of states of such a bilayer. Our
calculation is done both analytically and numerically in the
ballistic limit, using both a continuum- and lattice-model. It is
found that the spin-dependent phase-shifts occuring at the
HM$\mid$S interface seriously influence all of the aforementioned
phenomena. We explain our results in terms of Andreev reflection in the presence of a spin-active
interface, allowing for both spin-filtering and spin-mixing
processes. We demonstrate how the surface-bound states induced by
the anisotropy of the superconducting order parameter at the
HM$\mid$S interface are highly sensitive to these spin-dependent
processes. Our results can be directly tested experimentally using
STM-measurements and/or point-contact spectroscopy.
  \end{abstract}
%\pacs{...}

\maketitle

\section{Introduction}

In recent years, the physics of composite
superconductor$\mid$ferromagnet systems has been subject to
intense investigations. Apart from a
wealth of interesting effects to explore from a fundamental
physics point of view,  it is also hoped that the interplay
between the dissipationless current flow in superconductors
combined with the spin-polarization in ferromagnets will lead the
way to new applications in low-temperature nanotechnology.
\par The mutual influence of superconducting and ferromagnetic
elements in heterostructures has a long history, see Ref.
\cite{bergeret_rmp_05, buzdin_rmp_05} and references therein. While the
basic constituent in a superconducting condensate is a spin-singlet Cooper
pair in the usual Bardeen-Cooper-Schrieffer \cite{bardeen_pr_57} paradigm,
the superconducting correlations are strongly altered when placed
in close proximity to a ferromagnetic system, which spontaneously
breaks time-reversal symmetry. whenever translational
symmetry or time-reversal symmetry is broken, Cooper pairs with
unconventional pairing correlations are formed in general
\cite{berezinskii_pisma_74, eschrig_ltp_07, tanaka_prl_07}. Such
pairing correlations are unconventional in the sense that they
differ from the conventional spin-singlet Cooper pairs, and they
may exhibit for instance a spin-triplet symmetry or an
odd-frequency symmetry. The study of the proximity effect in
superconductor$\mid$ferromagnet heterostructures has received a
lot of attention in recent years (see, \textit{e.g.},
Refs.~\onlinecite{bergeret_prl_01, ryazanov_prl_01,
volkov_prl_03,bergeret_prb_03,eschrig_prl_03,braude_prl_07,
asano_prl_07_1,keizer_nature_06,fominov_prb_07,yokoyama_prb_07,
asano_prl_07_2,halterman_prl_07,tanaka_prl_07_2,eschrig_jlow_07,linder_prb_08,
eschrig_nphys_08,halterman_prb_08,linder_prb_08_2,bulaevskii_jetp_77,buzdin_pisma_82,
koshina_prb_01,kontos_prl_02,buzdin_prb_03,houzet_prb_05,cottet_prb_05,robinson_prl_06,
zareyan_prb_06,linder_prl_08,cuoco_prb_09,champel_prl_08,volkov_prb_08}).
\par
In the extreme ferromagnetic limit of a half-metal, where the
spin-polarization is close to 100\%, one would naively expect
proximity-induced superconducting correlations to be destroyed due
to the large exchange field in the ferromagnet. However, quite
surprisingly at the time, Keizer \textit{et al.} found
\cite{keizer_nature_06} that a supercurrent could flow between two
conventional $s$-wave superconductors separated by a half-metallic
layer of considerable size ($\sim\mu$m). This finding prompted
several authors to investigate the underlying physics that
permitted the superconducting correlations to survive over a long
distance in a fully polarized ferromagnet \cite{eschrig_nphys_08,
asano_prl_07_2}.
\par
Prior to the experimental finding in Ref. \cite{keizer_nature_06},
the concepts of \textit{spin-mixing} and \textit{spin-flip}
processes were drawn upon in Ref. \cite{eschrig_prl_03} in order
to explain how a supercurrent could be generated and sustained in
an $s$-wave/half-metal/$s$-wave junction. The scattering of
quasiparticles at the interface may in general be spin-dependent
in the presence of magnetic parts of the system, which is the case
for a superconductor/ferromagnet junction. This renders the
transmission probabilities for spin-$\uparrow$ and
spin-$\downarrow$ particles different not only in magnitude, but
also through the phases they pick up upon scattering at the
interface. This gives rise to a so-called spin-mixing at the
interface, which allows the singlet amplitude to be converted into
a $S_z=0$ triplet component, since scattered electrons with
opposite spins experience different phase shifts at the interface.
As a result, the superconducting correlations become a
superposition of both singlet and $S_z=0$ triplet pairing.  It is convenient for
later use to briefly recapitulate here how this happens \cite{eschrig_prl_03}.
Consider a singlet correlation function in the
superconductor:
\begin{align}\label{eq:singlet} \vert\psi\rangle &=
\vert\!\!\uparrow\rangle_k\vert\!\!\downarrow\rangle_{-k} -
\vert\!\!\downarrow\rangle_k\vert\!\!\uparrow\rangle_{-k}.
\end{align}
Upon scattering at the interface, the spins acquire different phase shifts
\begin{align}
\vert\!\!\uparrow\rangle_{-k} =
\e{\i\theta_\uparrow}\vert\!\!\uparrow\rangle_k,\;
\vert\!\!\downarrow\rangle_{-k} =
\e{\i\theta_\downarrow}|\downarrow\rangle_k.
\end{align}
This transforms Eq. (\ref{eq:singlet}) into
\begin{align}
\vert\psi\rangle &= -\cos(\Delta\theta)
\Big(\vert\!\!\uparrow\rangle_k\vert\!\!\downarrow\rangle_{-k} -
\vert\!\!\downarrow\rangle_k\vert\!\!\uparrow\rangle_{-k}\Big)
\notag\\
& -\i\sin(\Delta\theta)
\Big(\vert\!\!\uparrow\rangle_k\vert\!\!\downarrow\rangle_{-k} +
\vert\!\!\downarrow\rangle_k\vert\!\!\uparrow\rangle_{-k}\Big).
\end{align}
Here, $\Delta\theta = \theta_\uparrow-\theta_\downarrow$.
The spin-dependent phase-shifts at the interface induce a triplet
component which contributes to the total wavefunction
$\vert\psi\rangle$ as long as $\Delta\theta\neq0$.
\par However, it is also necessary to generate an equal-spin pairing
$S_z=\pm1$ components in order to sustain the long-range triplet
correlations. This demands spin-flip scattering processes of
the type $\vert\!\!\uparrow\rangle_k \to
\vert\!\!\downarrow\rangle_k$ and $\vert\!\!\downarrow\rangle_k
\to \vert\!\!\uparrow\rangle_k$ close to the interface. Such
processes are unavoidably present for instance in the case where
there are local inhomogeneities of the magnetic moment near the
interface. The combination of spin-mixing and spin-flip processes
then explain how the spin-singlet $s$-wave component of the bulk
superconductor may be converted into a long-range spin-triplet
component that is able to survive the large exchange field in the
half-metallic region.
\par
The above discussion underlines the crucial importance of treating
the interface properties correctly, and specifically taking into
account the spin-dependent phase-shifts that may occur for the
particles participating in the scattering processes
\cite{linder_prl_09}. In addition, the presence of mixed-parity
pairing correlations in a S$\mid$HM structure should be linked to
the spin-active nature of the interface. Previous literature has
considered only the proximity effect between conventional $s$-wave
superconductors and half-metallic ferromagnets
\cite{eschrig_nphys_08}. In the present work, our aim is to
investigate the interplay between the spin-dependent interface
properties and \textit{unconventional} pairing symmetries in the
bulk superconductor with regard to the \textit{(i)} conductance
spectra, \textit{(ii)} proximity effect, and \textit{(iii)} local
density of states of such a bilayer. These quantities are directly
accessible in experiments via STM-measurements and/or
point-contact spectroscopy. In particular, by allowing for an
unconventional pairing symmetry in the superconductor, such as
$p$-wave or $d$-wave, we may investigate the interplay between
Andreev-bound surface states \cite{hu_prl_94, tanaka_prl_95} and
half-metallicity.
\par We
organize this work as follows. In Sec. \ref{sec:theory}, we
present the theoretical formulation used in this work, namely the
Bogoliubov-de Gennes formalism. In Sec.
\ref{sec:results_conductance} and \ref{sec:results_DOS}, we
present and discuss our results for the conductance and proximity
effect/DOS, respectively. Finally, we give our conclusions in Sec.
\ref{sec:summary}. We will use boldface notation for 3-vectors,
$\hat{\ldots}$ for $4\times4$ matrices, and $\underline{\ldots}$
for $2\times2$ matrices.

\section{Theory}\label{sec:theory}

\begin{figure}[b!]
\centering \resizebox{0.30\textwidth}{!}{
\includegraphics{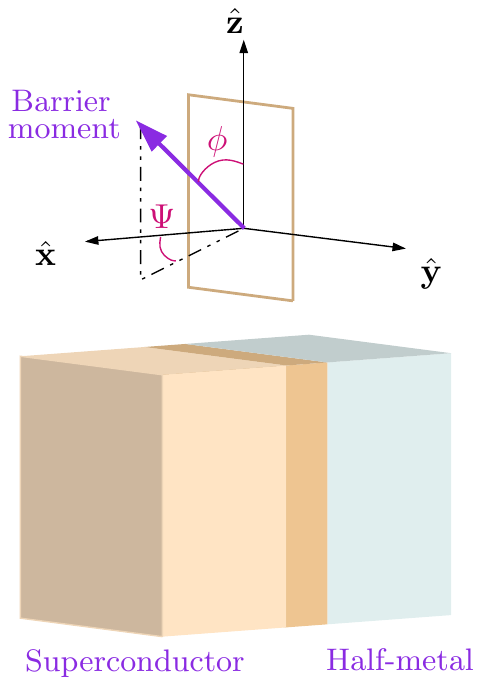}}
\caption{(Color online) The superconductor/half-metallic
ferromagnet bilayer studied in this paper. The barrier magnetic
moment may in general be misaligned to the bulk magnetization in
the ferromagnet, which is assumed to be directed along
$\mathbf{\hat{z}}$. The presence of a barrier magnetic moment may
lead to both spin-split and spin-flip processes at the
interface.}\label{fig:model_conductance}
\end{figure}

In order to calculate the conductance of the S/HM junction, we
apply a modified Blonder-Tinkham-Klapwijk (BTK) theory which takes
into account both an arbitrary pairing symmetry of the
superconductor as well as spin-mixing at the interface.
Specifically, we consider the situation as shown in Fig.
\ref{fig:model_conductance}, where the region near the interface
is allowed to have misaligned magnetic moments as compared to the
bulk of the half-metallic ferromagnet. Our starting point is the
BdG-equation
\begin{align} \hat{H}\Psi = \varepsilon\Psi
\end{align}
 in the half-metallic and superconducting region. We find that
\begin{widetext}
\begin{align}\label{eq:bdg}
\hat{H} = \begin{pmatrix}
H_0 - h_z\Theta(-x) + V_\uparrow\delta(x) & (V_x-\i V_y)\delta(x) & 0 & \Delta(\theta)\Theta(x) \\
(V_x+\i V_y)\delta(x) & H_0  + h_z\Theta(-x) + V_\downarrow\delta(x) & \zeta \Delta(\theta)\Theta(x) & 0 \\
0 & \zeta\Delta(\theta)^*\Theta(x) & -H_0   + h_z\Theta(-x) - V_\uparrow\delta(x) & -(V_x+\i V_y)\delta(x) \\
\Delta(\theta)^*\Theta(x) & 0  & -(V_x-\i V_y)\delta(x) & -H_0  -h_z\Theta(-x)- V_\downarrow\delta(x)\\
\end{pmatrix}
\end{align}
\end{widetext}
upon defining
\begin{align}
H_0=-\frac{\nabla^2}{2m} -\mu,\; V_\sigma = V_0+\sigma V_z,
\end{align}
while $\Theta(x)$ and $\delta(x)$ are the Heaviside step-function
and delta-function, respectively. Here, the barrier magnetic
moment constitutes a spin-dependent potential, where $V_x = -\rho
V_0\cos\Psi\sin\phi$, $V_y=-\rho V_0\sin\Psi\sin\phi$, $V_z =
-\rho V_0\cos\phi$. The intrinsic non-magnetic barrier potential
is $V_0$, while $\rho$ constitutes the effective ratio between the
non-magnetic and magnetic barrier, since
\begin{align} \rho =
|\mathbf{V}|/V_0
\end{align}
where $\mathbf{V} = (V_x,V_y,V_z)$. The parameter $\zeta$ accounts
for singlet or triplet pairing through $\zeta=-1$ for singlet
pairing while $\zeta=1$ for triplet pairing. In both cases,
however, we assume opposite-spin pairing, corresponding to a
unitary state in the triplet case. The exchange energy in the
half-metallic ferromagnet is modelled through $h_z$, and we will
later take the limit $h_z\to\mu$, corresponding to a fully
polarized ferromagnet.
\par Solving Eq. (\ref{eq:bdg}), we obtain
the following wavefunction in the superconducting region:
\begin{align}\label{eq:psiS}
\psi_S(x) &=  t_e^\uparrow\Big[u(\theta_S),0,0,v(\theta_S)\e{-\i\gamma_+}\Big]\e{\i q\cos\theta_Sx}\notag\\
&+ t_e^\downarrow\Big[0,u(\theta_s),\zeta v(\theta_s)\e{-\i\gamma_+},0\Big]\e{\i q\cos\theta_Sx}\notag\\
&+ t_h^\uparrow\Big[0,\zeta v(\pi-\theta_S)\e{\i\gamma_-}, u(\pi-\theta_S),0\Big]\e{-\i q\cos\theta_Sx}\notag\\
&+ t_h^\downarrow\Big[v(\pi-\theta_S)\e{\i\gamma_-},0,0,u(\pi-\theta_S)\Big]\e{-\i q\cos\theta_Sx},
\end{align}
while in the ferromagnetic region we have for an incoming
spin-$\uparrow$ electron with positive excitation energy
$\varepsilon$:
\begin{align}\label{eq:psiHM}
\psi_{HM}(x)&= \Big(\e{\i k^\uparrow\cos\theta x} + r_e^\uparrow\e{-\i k^\uparrow\cos\theta x}\Big) \Big[1,0,0,0\Big]\notag\\
&+ r_e^\downarrow\Big[0,1,0,0\Big]\e{-\i k^\downarrow\cos\theta^\downarrow x} \notag\\
&+ r_h^\uparrow\Big[0,0,1,0\Big]\e{\i k^\uparrow \cos\theta_A^\uparrow}\notag\\
&+ r_h^\downarrow\Big[0,0,0,1\Big]\e{\i k_A^\downarrow \cos\theta_A^\downarrow}.
\end{align}
In the above equations, $\{t_e^\sigma, t_h^\sigma\}$ denote the
transmission coefficients for electron-like and hole-like
quasiparticles in the superconductor with spin $\sigma$. Note that
without any spin-mixing at the interface, one has
$t_e^\downarrow=t_h^\uparrow=0$ in the present case. This is
because that an incoming spin-$\uparrow$ from the HM side can only
be reflected normally in such a scenario without conversion at the
interface. We comment more on this later. Moreover, $q$ denotes
the Fermi-level momentum in the S region while $\theta_S$ is the
propagation angle. The coherence functions are defined in the
standard way:
\begin{align}
u(\theta) &= \sqrt{\frac{1}{2}\Big(1 + \frac{\sqrt{\varepsilon^2 - |\Delta(\theta)|^2}}{\varepsilon}\Big)},\notag\\
v(\theta) &= \sqrt{\frac{1}{2}\Big(1 - \frac{\sqrt{\varepsilon^2 - |\Delta(\theta)|^2}}{\varepsilon}\Big)}
\end{align}
We have also introduced the phase factors
\begin{align}
\e{\i\gamma_\pm} = \Delta(\theta_\pm)/|\Delta(\theta_\pm)| \text{ with } \theta_+=\theta,\; \theta_-=\pi-\theta.
\end{align}
In the normal region, we have artificially included
scattering-coefficients for spin-$\downarrow$ since our strategy
is to do the calculation for a strong ferromagnet, and finally
take the half-metallic limit, corresponding to $\{k^\downarrow,
k_A^\downarrow\}\to0$. In the final expression for the
conductance, the contribution from minority spin will be down a
factor $|k^\uparrow/k^\downarrow|\to\infty$ compared to the
majority spin contribution, which gives us the correct result in
the half-metallic limit. We choose $q=k^\uparrow$, assuming that
Fermi-vector mismatch effects simply alter the effective barrier
resistance. In this case, $\theta_S=\theta$.
\par The task at hand
is now to calculate the scattering coefficients, which are needed
to evaluate the conductance. To do so, we need to incorporate
proper boundary conditions. The presence of a magnetic moment in
the barrier, which is not necessarily aligned with the
magnetization in the bulk HM region, introduces new components in
the boundary conditions as compared to the ones that mostly have
been used in the literature. Assuming a barrier with a
spin-independent potential $V_0$ and a spin-dependent potential of
strength $\rho V_0$, where the orientation of the magnetic moment
is  described by two angles $\{\phi,\Psi\}$ as shown in Fig.
\ref{fig:model_conductance}, we may write:
\begin{align}\label{eq:boundary}
\partial_x[\psi_S(x)-\psi_{HM}(x)]|_{x=0} &= 2mV_0[\hat{1} - \rho\cos\phi (\underline{\tau}_0\otimes\underline{\sigma_3}) \notag\\
&- \rho\sin\phi \hat{M}(\Psi)]\psi_{HM}(0),
\end{align}
where we have defined the matrices
\begin{align}
\hat{M}(\Psi) &= \begin{pmatrix}
\underline{\Theta}(\Psi) & 0 \\
0 & \underline{\Theta}^*(\Psi) \\
\end{pmatrix},\; \underline{\Theta}(\Psi) =\begin{pmatrix}
0 & \e{-\i\Psi}\\
\e{\i\Psi} & 0 \\
\end{pmatrix}
\end{align}
In addition, continuity of the wavefunction gives $\psi_{HM}(0) =
\psi_S(0)$. Using these boundary conditions with Eqs.
(\ref{eq:psiS}) and (\ref{eq:psiHM}), one obtains the solution for
the scattering coefficients. Then, the conductance of the junction
is expressed at zero temperature through the dimensionless
quantity
\begin{align} G(eV) = \int^{\pi/2}_{-\pi/2} \text{d}\theta
f(\theta)[1 + |r_h^\uparrow(eV)|^2 - |r_e^\uparrow(eV)|^2.
\end{align}
Here, $f(\theta)$ is a angle-dependent factor which models the
probability distribution for incoming electrons at an angle
$\theta$. Usually, it is chosen to $f(\theta)=\cos\theta$ to favor
angles close to normal incidence, but it may also be chosen to
exhibit a more discriminating tunneling cone behavior. In all the
plots, we will normalize the conductance on its value at voltages
much larger than the gap, i.e. $eV\gg\Delta_0$, as is usually done
when comparing against experimental data since this regime
corresponds to the normal-state conductance, and we choose
$f(\theta)=\cos\theta$.
\par Although an analytical solution for
the scattering coefficients is possible in principle, the
resulting expressions are somewhat cumbersome, so we omit them
here. For a fixed pairing symmetry $\Delta(\theta)$, the interface
properties will determine the behavior of the junction
conductance. The interface parameters are then the
spin-independent barrier strength $Z=2mV_0/q$, the ratio between
the spin-independent and spin-dependent scattering potential
$\rho$, and the orientation of the barrier magnetic moment
determined by $\phi$ and $\Psi$. All of these quantities are
dimensionless.
\par
In what follows, we will fix the barrier strength at $Z=3$
corresponding to a weakly transparent interface, since this should
correspond to a realistic experimental situation. All the
interesting physics then lies in the parameters $\{\rho,\phi,
\Psi\}$. In the experimental work so far, strong sample-to-sample
variations are seen in the results for the conductance and the
critical current of S/HM heterostructures. As pointed out in Ref.
\cite{eschrig_nphys_08}, this is an indication that the
spin-properties of the interface vary greatly between different
samples, suggesting that the quantities $\{\rho,\phi, \Psi\}$ are
very hard to control experimentally. The purpose of this paper is to
obtain a fuller picture of how the spin-active properties of the
interface influence the conductance spectra in order to gain a
clearer understanding of the characteristic features seen in the
experimental data.

\section{Results: Conductance}\label{sec:results_conductance}

Before we proceed to a dissemination of our results, let us
establish contact between the terminology and notation used in
previous literature regarding proximity structures of
superconductors and half-metallic ferromagnets. In our notation,
spin-mixing corresponds to the case of a finite $\rho$, but with
$\phi=0$. In this case, the effective potential felt by
spin-$\uparrow$ and spin-$\downarrow$ electrons scattered at the
interface is different, and they pick up different phases along
their scattering trajectories. Note that the spin-dependent
potential also gives rise to, in general, a magnetoresistance
effect known as spin-filtering since the transmission amplitudes
for opposite spins are not the same. Spin-flip scattering,
however, requires a misalignment between the barrier magnetic
moment and the bulk magnetization in the half-metallic
ferromagnet. In our notation, it is then necessary to have
$\rho\neq0$ and also $\phi\neq0$. Only then will the scattering
amplitudes $t_e^\downarrow$ and $t_h^\uparrow$ be non-zero in
general, as commented on earlier. From Fig.
\ref{fig:model_conductance}, it is clear that it suffices to vary
only $\phi$ in order to obtain both spin-mixing and spin-flip
processes. To reduce the number of free parameters and still grasp
the key physics, we therefore set $\Psi=\pi/2$ in what follows. In
this way, the barrier magnetic moment lies in the $y-z$ plane.
Spin-mixing is then obtained for $\rho\neq0$ and $\phi=0$, while
spin-mixing \textit{and} spin-flip processes are obtained for
$\rho\neq0$ and $\phi\neq0$.
\par So far in the literature, the
interplay between unconventional bulk superconductivity and
half-metals has not been studied yet. We will therefore consider
several bulk pairing symmetries in the superconducting region,
including $p$- and $d$-wave pairing.

\subsection{$s$-wave pairing}

\begin{figure*}
\centering
\resizebox{0.9\textwidth}{!}{
\includegraphics{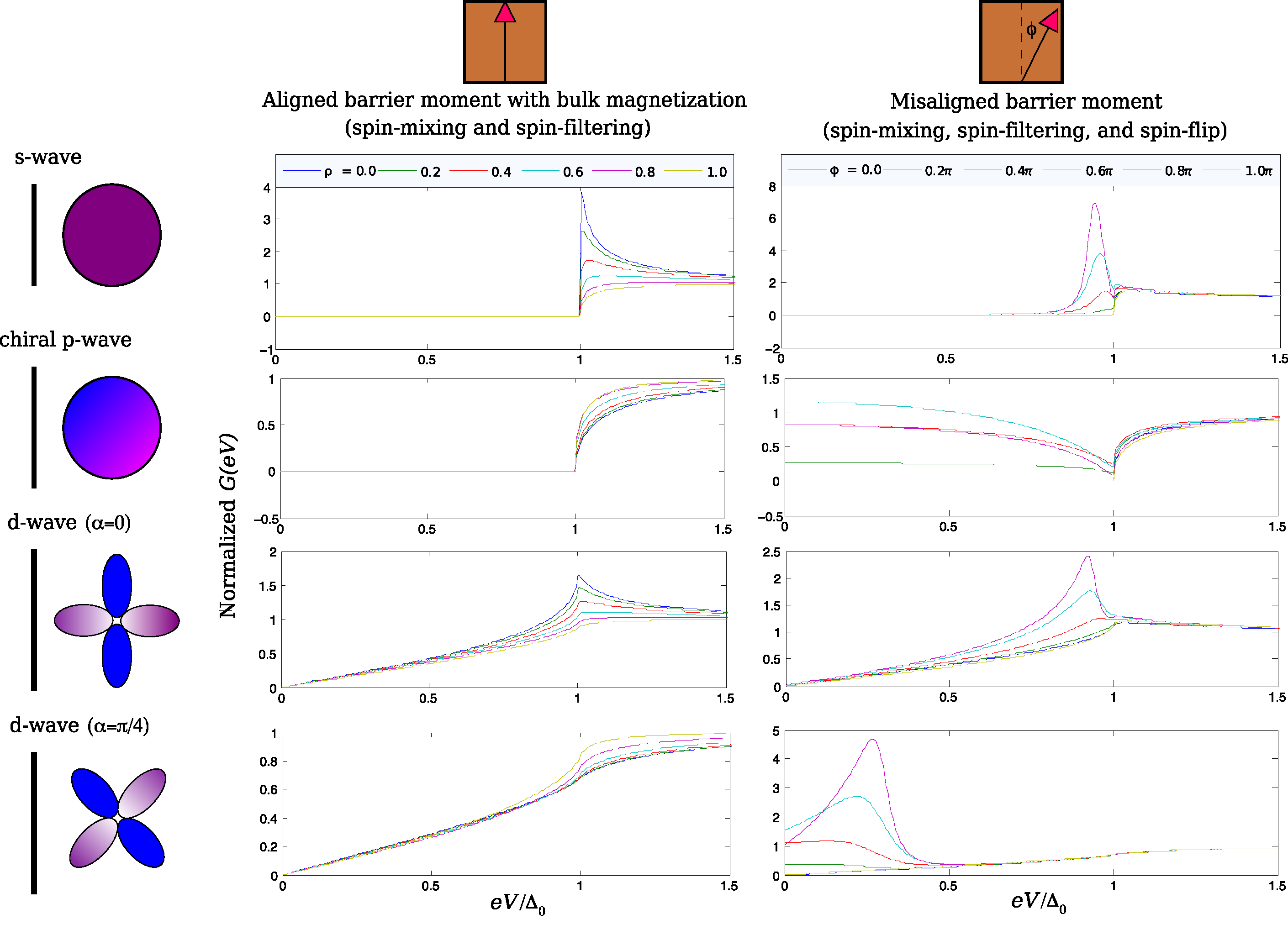}}
\caption{(Color online) Plot of the normalized conductance $G(eV)$
for a $s$-wave, chiral $p$-wave, $d_{x^2-y^2}$-wave, and
$d_{xy}$-wave symmetry in the rows ranging from top to bottom. In
the left panels, we consider the case of pure spin-mixing for
several values of $\rho$ with $\phi=0$. In the right panels, we
consider additionally spin-flip processes induced by a misaligned
barrier moment at the interface for several values of $\phi$ with
$\rho=0.5$.}\label{fig:cond}
\end{figure*}

For $s$-wave pairing, we choose $\Delta(\theta)=\Delta_0$.
Consider first the situation of pure spin-mixing, corresponding to
$\rho\neq0$, while $\phi=0$. This is shown in the first row of
Fig. \ref{fig:cond} (left panel), where it is seen that the sharp
coherence peak at the gap is replaced with broadened features upon
increasing $\rho$. This is in agreement with Fig. 2 of Ref.
\cite{kopu_prb_03} for high values of their parameter $R$, which
corresponds roughly to our $Z$. Note that the subgap conductance
is exactly zero, regardless of the value of $\rho$. The reason for
this is that the usual Andreev-reflection where the hole has
opposite spin of the incoming electron is not possible in the
present case of a half-metal, unless spin-flip processes are
allowed at the interface. Interestingly, Andreev-reflection is
therefore absent in the system regardless, in fact, of the value
of $Z$ unless there is a magnetically inhomogeneous region near
the interface. We will discuss this on a microscopic level in much
more detail in Sec. \ref{sec:analytical}.
\par To illustrate how
spin-flip processes affect the conductance, we show in Fig.
\ref{fig:cond} (right panel) the case of $\rho=0.5$ for several
values of $\phi$. As seen, once $\phi$ becomes non-zero, the
subgap conductance becomes finite. A large peak very close to
$eV=\Delta_0$ evolves with increasing $\phi$. However, the
zero-bias conductance remains suppressed regardless of the
orientation of the barrier moment. In Ref.
\cite{krivoruchko_prb_08}, the conductance was experimentally
measured for a Pb/La$_{0.7}$Sr$_0.3$MnO$_3$ point contact setup,
where the authors found strong sample-to-sample variations. Some
of the samples showed clear zero-bias conductance peaks, which is
usually a signature of odd-frequency correlations or zero-energy
Andreev-bound states. Other samples displayed a clear minigap
structure similar to our finding in Fig. \ref{fig:cond} for the
$s$-wave case. The strong zero-bias peak observed in the samples
led the authors of Ref. \cite{krivoruchko_prb_08} to speculate
that an even-frequency $p$-wave bulk state was induced in
half-metallic La$_{0.7}$Sr$_{0.3}$MnO$_3$ by means of the
proximity to superconducting Pb, thus rendering the
Pb/La$_{0.7}$Sr$_{0.3}$MnO$_3$ junction into an S/S junction.
Another observation in Ref. \cite{krivoruchko_prb_08} that
supported this idea was a spectacular drop of the contact's
resistance with the onset of the Pb superconductivity.

\subsection{$p$-wave pairing}
For $p$-wave pairing, we will consider a gap of the form
$\Delta(\theta)=\Delta_0\e{\i\theta}$. This is known as chiral
$p$-wave or $p_x+ip_y$-wave pairing, and is believed to be
realized in Sr$_2$RuO$_4$. The gap supports Andreev-bound
zero-energy states at normal incidence of incoming quasiparticles,
$\theta=0$, where it satisfies the appropriate symmetry condition
$\Delta(\theta) = -\Delta(\pi-\theta)$. The situation changes,
however, in the present case where the non-superconducting region
is half-metallic. Let us first consider the case of pure
spin-mixing $(\phi=0)$ in the second row of Fig. \ref{fig:cond}
(left panel). As seen, the subgap conductance is still zero, since
the gap magnitude $|\Delta(\theta)|$ is isotropic and thus
prevents direct quasiparticle tunnelling into any nodes of the
gap. Also, the usual zero-energy Andreev-bound states do not take
part in the scattering processes since there is no possibility for
Andreev-reflection of a spin-$\downarrow$ hole in the half-metal.

The situation changes drastically once we introduce magnetic
inhomogeneities at the interface (right panel), corresponding to
$\phi\neq0$. The subgap conductance, in particular the zero-bias
conductance, is greatly enhanced upon increasing $\phi$. The
reason is that although no spin-$\downarrow$ holes are available
in the half-metal, the presence of spin-flip scattering when
$\phi\neq0$ allows for Andreev reflection with spin-$\uparrow$
holes. The presence of Andreev reflection of majority spin holes
is the reason for the enhancement of the conductance.

\subsection{$d$-wave pairing}
For $d$-wave pairing, we choose
$\Delta(\theta)=\Delta_0\cos(2\theta-2\alpha)$. For $\alpha=0$,
there are no Andreev-bound states, while for $\alpha=\pi/4$ the
order parameter supports the formation of Andreev-bound
states in a N$\mid$$d$-wave junction. Consider first the case with
only spin-mixing at the interface, i.e. $\phi=0$. The
existence of nodes in the gap renders the subgap conductance
non-zero for both crystallographic orientations $\alpha=0$ and
$\alpha=\pi/4$. The effect of increasing
$\rho$ is opposite for the two orientations. For $\alpha=0$, the
conductance evolves from the typical $d$-wave bulk density of
states profile at $\rho=0$ to exhibit broader features at
$\rho\simeq1$. For $\alpha=\pi/4$ the conductance
evolves from broad features at $\rho=0$ to a typical $d$-wave bulk
density of states at $\rho\simeq1$.

Introducing spin-flip processes at the interface by allowing
$\phi\neq0$, the distinction between the two crystallographic
orientations becomes clear. For $\alpha=0$, the conductance is
similar to the bulk density of states, while for $\alpha=\pi/4$
the zero-bias conductance is strongly enhanced upon increasing
$\phi$. For the same reason as described in the chiral $p$-wave
case, this enhancement is a result of the spin-flip induced
Andreev reflection of spin-$\uparrow$ holes made possible by
$\phi\neq0$. Interestingly, a large peak evolves at an energy
inside the gap similarly to the $s$-wave case. One may thus
ask whether the presence of a subgap peak in the
conductance is indicative of surface bound-states induced by the
spin-active interface. We shall discuss this question in more
detail in the following section.

\subsection{Analytical expressions and bound-states}\label{sec:analytical}

In order to understand further the above results, it is
instructive to consider analytically the expression for the
scattering coefficients. In particular, we focus on the
Andreev reflection probability $r_h^\uparrow$ which only exists
for $\uparrow$-spin in the HM region. We find that the following
general expression:
\begin{align}\label{eq:andreev}
r_h^\uparrow &= -4\rho ZR^{-1}\sin\phi\cos^2\theta \e{\i\beta-\i\gamma_+} \Big[\i Z(1+\rho\cos\phi)(1+\zeta)\notag\\
&\times(\e{-\i\Delta\gamma}-\e{2\i\beta}) - \cos\theta(1-\zeta)(\e{-\i\Delta\gamma}+\e{2\i\beta}) \Big],
\end{align}
where we have defined $\e{\i\beta} = u(\theta)/v(\theta)$
and
\begin{align}
R &= Z^4(1-\rho^2)^2(\e{2\i\beta} - \e{-\i\Delta\gamma})^2 + 4\cos^4\theta\e{4\i\beta} + Z^2\cos^2\theta\notag\\
&\times \Big[\e{-2\i\Delta\gamma}(1-\rho\cos\phi)^2 + \e{2\i\beta-\i\Delta\gamma}[4\rho(\rho - \cos\phi) \notag\\
&- (1-\cos^2\phi)] + \e{4\i\beta}(6\rho\cos\phi + 4\rho^2+\rho^2\cos^2\phi+5)\Big].
\end{align}
Note that $r_h^\uparrow$ vanishes when either $\phi$, $\rho$, or
$Z$ are equal to zero. This is physically reasonable, since the
interface becomes spin-inactive in all those cases. Hence,
there are no spin-flip processes which can mediate Andreev
reflection $r_h^\uparrow$. Consider now the triplet pairing case
$\zeta=1$. In a normal metal$\mid$chiral $p$-wave superconductor
junction, it is well-known that the bound-state energies at the
interface have a dispersion $\varepsilon \sim \Delta_0\sin\theta$.
More specifically, the bound-state condition is given by
\begin{align}
2\beta = -\Delta\gamma = \pi-2\theta.
\end{align}
Interestingly, the Andreev-reflection coefficient Eq.
(\ref{eq:andreev}) vanishes completely for precisely these
energies, regardless of the other parameters in the system. This
is then opposite to the N$\mid$chiral $p$-wave case where the
Andreev reflection coefficient is unity at the bound-state
energies.

\begin{figure}[t!]
\centering
\resizebox{0.5\textwidth}{!}{
\includegraphics{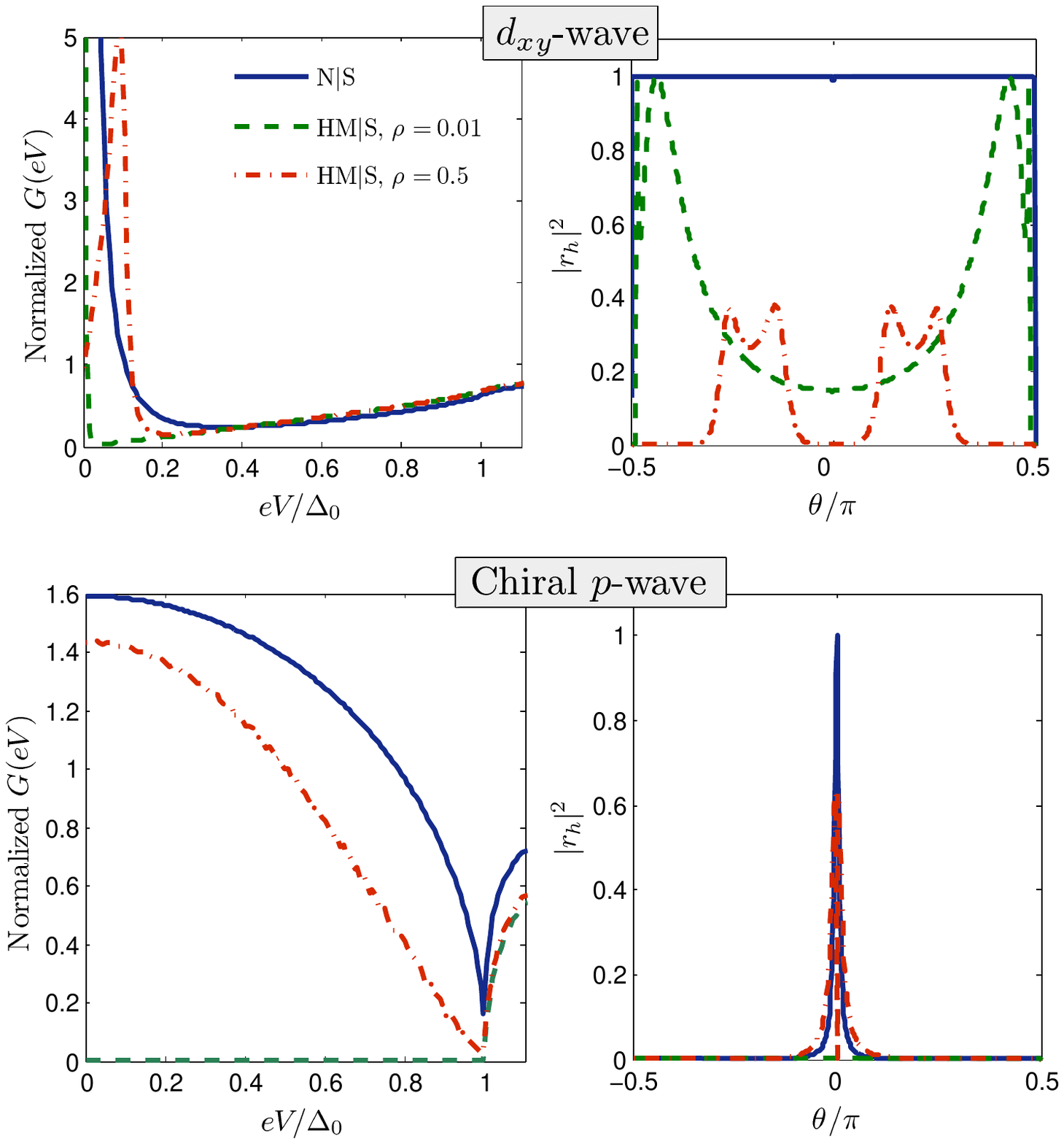}}
\caption{(Color online) Comparison of the conductance spectra for
N$\mid$S and HM$\mid$S junctions with a $d_{xy}$-wave
superconductor in the top row and a chiral $p$-wave superconductor
in the bottom row. In the half-metallic case, we consider a
spin-active interface with a misorientation angle $\phi=0.5$. The
ratio between the magnetic and non-magnetic part of the barrier
potential is denoted $\rho$. In all cases, the strong tunneling
limit $Z=10$ is considered. The Andreev reflection probability in
the right panels is calculated at the peak energy in the
$d_{xy}$-wave case while it is calculated at $\varepsilon=0$ in
the chiral $p$-wave case.}
\label{fig:pdwave}
\end{figure}

In the previous section, we pointed to the possibility that the
emergence of strong peaks in the subgap conductance seen in both
the $s$- and $d_{xy}$-wave cases is a signature of surface
bound-states. This would be similar to the zero-bias conductance
peak in N$\mid$$d_{xy}$-wave junctions originating from the
existence of zero-energy surface states. It should be noted that
an enhancement of the conductance above its normal-state value is
in general not sufficient to prove the existence of surface
bound-states. To see this, consider \eg a N$\mid$$s$-wave junction
with a good interface contact $(Z\leq 1)$, where Andreev
reflection occurs with a high probability even without any
interface bound-states. To clarify whether an enhancement of the
conductance (such as a resonant peak-structure) truly pertains to
surface bound-states, one has to consider the tunneling limit of a
strong barrier potential, or equivalently a low interface
transparency. If the conductance is still enhanced compared to its
normal-state value due to the presence of Andreev reflection, it
\textit{could} be a signature of resonant tunneling into a
surface-state. We now compare the behavior of the Andreev
reflection probability in the half-metallic limit with the
corresponding non-magnetic case in order to acquire information
about the origin of the conductance peak. In the top row of Fig.
\ref{fig:pdwave}, we plot the conductance of a
N$\mid$$d_{xy}$-wave junction (without a spin-active interface)
and a HM$\mid$$d_{xy}$-wave junction as well as the respective
Andreev reflection probabilities at the peak energies. The Andreev
reflection probability $|r_h|^2$ is given as a function of the
angle of incidence $\theta$. For all plots, we have set $Z=10$,
corresponding to strong tunneling limit.

In the normal metal case, the usual ZBCP is recovered and the
Andreev reflection probability is unity for all angles of
incidence. Thus, charge is transmitted into
the superconductor as a Cooper pair via the resonant zero-energy
states. Turning to the HM case, we wish to distinguish between the
two cases of a weak and strong magnetic moment of the barrier. For
$\rho=0.01$, it is seen that the ZBCP remains, while for
$\rho=0.5$ the ZBCP is shifted to a finite bias voltage. In the
latter case, similar behavior was also reported for a
N$\mid$$d_{xy}$-wave junction with a spin-active interface in Ref.
\cite{kashiwaya_prb_99}. In our case, it is necessary to have a
non-zero misalignment angle $\phi$ between the barrier moment and
the bulk magnetization in order to generate Andreev reflection at
all, contrary to the scenario of Ref. \cite{kashiwaya_prb_99}.
However, the Andreev reflection coefficients shown for the
$d_{xy}$-wave case in Fig. \ref{fig:pdwave} indicate that the
peaks cannot be ascribed to resonant energy states that are
available at all angles of incidence. In fact, the probability for
Andreev reflection never reaches unity when $\rho=0.5$. Still,
$|r_h^\uparrow|^2$ is substantial in magnitude even though we are
considering the tunneling limit. It therefore appears that
surface-states are induced close to the interface which enable
transmission processes in spite of the large barrier potential,
although they are not resonant in the sense that transmission into
them occurs with a probability of unity.

Turning now to the chiral $p$-wave case in the lower row of Fig.
\ref{fig:pdwave}, we see that the subgap conductance remains
completely suppressed when $\rho$ is small. This contrasts with the
$d_{xy}$-wave case. We next increase the magnitude of the magnetic part
of the barrier compared to the non-magnetic part further, i.e.
we increase $\rho$. It is now seen that the subgap conductance then
becomes comparable to the N$\mid$chiral $p$-wave case. Note that both
spin species see an effective barrier potential in
the tunnelling limit even for $\rho=0.5$ due to the large value of
$Z$. Therefore, the large enhancement of the subgap conductance
must stem from surface-induced states which decay inside the bulk.
In the following section, we will employ a self-consistent
Bogoliubov-de Gennes framework to numerically investigate whether
the local DOS near the interface truly features such surface-bound
states or not.
\begin{figure*}
\centering
\resizebox{0.8\textwidth}{!}{
\includegraphics{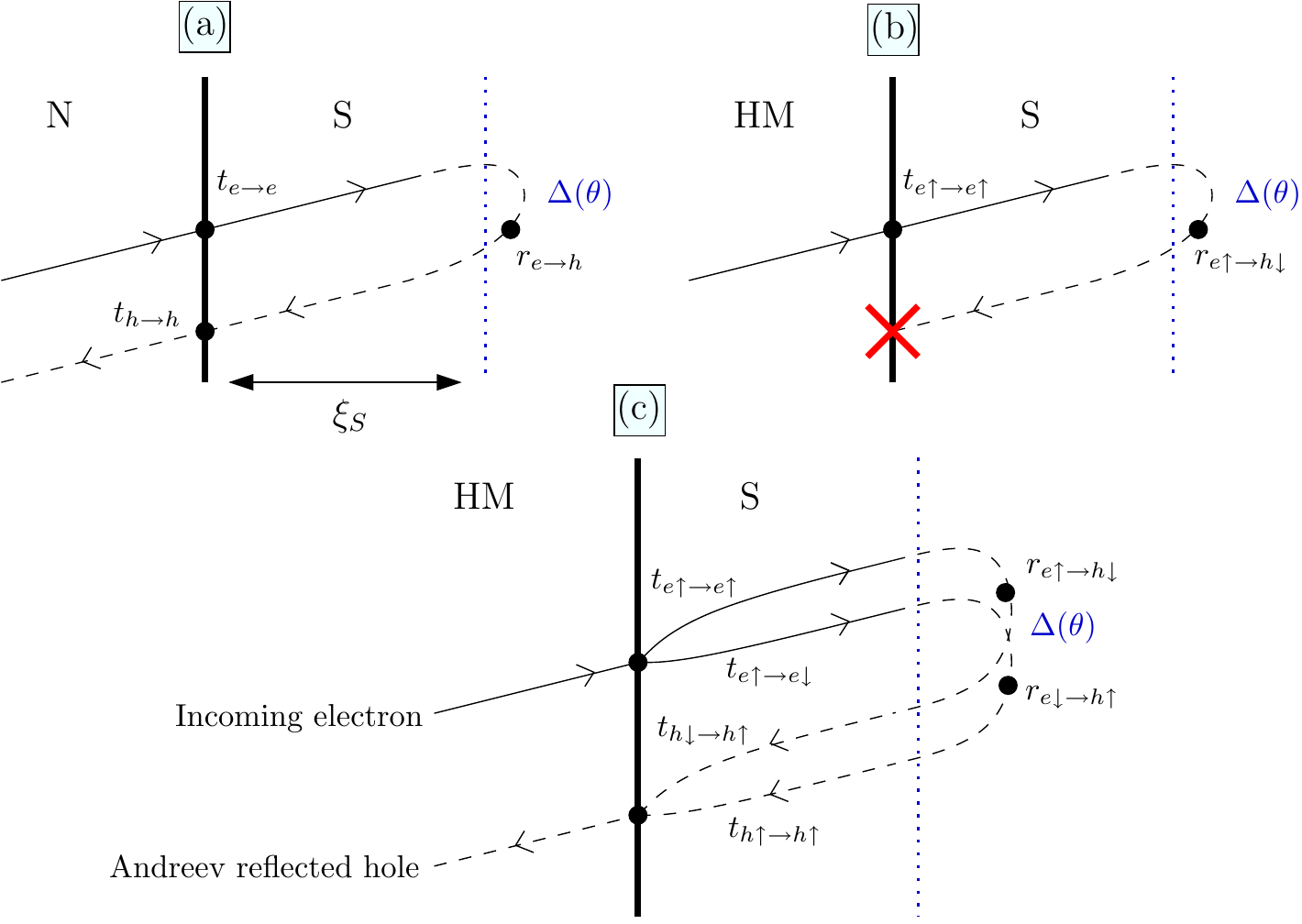}}
\caption{(Color online) Andreev reflection scattering to first
order for three cases: (a) N$\mid$S junction without spin-active
interface, (b) HM$\mid$S junction without spin-active interface,
(c) HM$\mid$S junction with spin-active interface. The incoming
electron-like quasiparticle penetrates the superconducting region
a distance $\xi_S$ before being backscattered as a hole by the
superconducting gap.}\label{fig:scattering}
\end{figure*}

To understand how the spin-active interface influences Andreev
reflection and midgap bound states on a microscopic level, it is
useful to again compare the N$\mid$S case with the HM$\mid$S case.
The Andreev reflection process to first order is shown in Fig.
\ref{fig:scattering}. Higher order processes may be generalized
from the first order process along the lines of Ref.
\cite{asano_prb_04}. From Fig. \ref{fig:scattering}, the general
recipe for Andreev reflection is seen to be a transmission of an
electron-like quasiparticle into the superconductor, which
penetrates about a coherence length $\xi_S$ before it is scattered
back as a hole by the gap $\Delta(\theta)$. Finally, the hole-like
quasiparticle is transmitted to the non-superconducting region. In
Fig. \ref{fig:scattering}(a), the Andreev reflection coefficient
to first order is thus seen to be $r_h^{(1)} = t_{h\to h} r_{e\to
h} t_{e\to e}$, and the contribution from higher order processes
is built along the same lines. We have omitted spin indices since
the spin for each process is uniquely defined: transmission
preserves spin while Andreev reflection flips spin. In Fig.
\ref{fig:scattering}(b), we consider a HM$\mid$S junction without
a spin-flip processes at the interface, corresponding to $\phi=0$.
As seen, Andreev reflection is rendered
impossible since the spin-$\downarrow$ hole backscattered by the
gap $\Delta(\theta)$ cannot be transmitted into the half-metallic
region due to the vanishing DOS for minority spin there. We restrict our
attention to opposite spin-pairing superconductors only, such as $s$-wave,
chiral $p$-wave ($S_z=0$), and $d$-wave. For an equal spin-pairing
superconductor, Andreev reflection is obviously possible even
without any spin-flip processes at the interface. In Fig.
\ref{fig:scattering}(c), we consider a HM$\mid$S junction with
spin-flip processes at the interface. As mentioned previously,
Andreev reflection is now possible even for a backscattered
minority spin hole due to the spin-flip probability at the
interface.

The microscopic picture shown in Fig. \ref{fig:scattering} also
allows us to understand how the midgap bound states are influenced
by the spin-active interface. To do so, we first briefly
recapitulate the results of Ref. \cite{asano_prb_04} for the
N$\mid$S case without spin-active processes. Focusing on
$\varepsilon=0$ where the retroreflection property holds in the S
region, one can calculate the total probability for Andreev
reflection obtained by summing all orders of scattering diagrams
such as the ones shown in Fig. \ref{fig:scattering}. In doing so,
the total probability is proportional to $\mid \sum_{n=0}^\infty
|r|^{2n}[-\e{\i(\gamma_--\gamma_+)}]^n\mid^2$, where $\gamma_\pm$
represents the phase contribution from the internal phase of the
superconducting order parameter while $|r| = |r_{e\to e}| =
|r_{h\to h}|$. For a $d_{xy}$-wave superconductor where
$\e{\i(\gamma_--\gamma_+)}=-1$, it is seen that all orders sum in
a coherent way and the total Andreev reflection probability can be
shown to equal unity. In the present case, the phases picked up by
the scattered particles are longer spin-degenerate. In particular,
we see from Fig. \ref{fig:scattering}(c) that the probability for
Andreev reflection to first order is equal to
\begin{align}
r_h^{(1)} = t_{e\uparrow\to e\uparrow} r_{e\uparrow\to
h\downarrow} t_{h\downarrow\to h\uparrow} + t_{e\uparrow\to
e\downarrow} r_{e\downarrow\to h\uparrow} t_{h\uparrow\to
h\uparrow} .
\end{align}
The crucial point is now that, whereas the
branch-converting reflection coefficients $r_{e\to h}$ have
spin-independent scalar phases without a spin-active interface,
they are spin-dependent otherwise. In the former case, one has
$r_{e\sigma\to h,-\sigma} \sim \e{-\i\gamma(\theta) +
\i\vartheta}$ and $r_{h\sigma\to e,-\sigma} \sim
\e{\i\gamma(\theta) - \i\vartheta}$ with
$\sigma=\uparrow,\downarrow$. When summing the Andreev reflection
processes to all orders, one obtains products of $r_{e\sigma\to
h,-\sigma}$ and $r_{h\sigma\to e,-\sigma}$ which effectively gives
a phase-factor $\e{\i(\gamma_--\gamma_+)}$ while the other scalar
phases cancel each other. When the interface is spin-active, the
scalar phases $\vartheta$ become spin-dependent, and one
effectively gets an additional contribution
$\Delta\vartheta_\sigma = \vartheta_\sigma - \vartheta_{-\sigma}$
in the phase of the effective Andreev reflection coefficient. For
this reason, the summation over all orders $n$ is altered and the
resonant states at \eg $\varepsilon=0$ in the $d_{xy}$-wave case
are shifted. The spin-dependent phase-shifts $\vartheta_\sigma$
depend on both $\rho$ and $\phi$ and are the reason for why the
conductance is qualitatively altered in the presence of a
spin-active interface as shown in both Fig. \ref{fig:cond} and
Fig. \ref{fig:pdwave}.

\section{Proximity effect and local density of states: a lattice study}\label{sec:results_DOS}

\begin{figure}[b!]
\centering \resizebox{0.50\textwidth}{!}{
\includegraphics{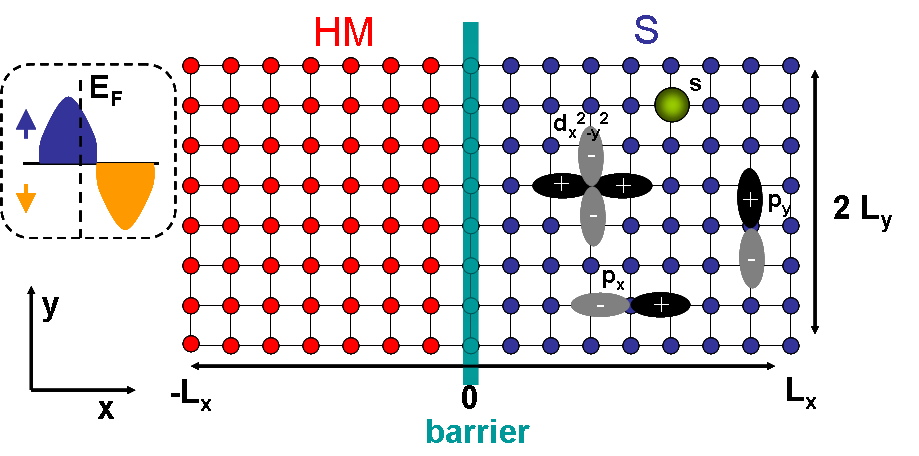}}
\caption{(Color online) Schematic description of the lattice geometry for the HM$\mid$S bilayer junction
indicating the reference axis system, the notation for the size (2$L_x\times$ 2$L_y$), the position of
the barrier ($x=0)$, the sketch of the pairing configurations (d-wave, p-wave and local s-wave), respectively.
On the left side, we show a sketch of the density of states for the half-metal. }\label{fig:lattice}
\end{figure}

In this section, within a self-consistent scheme of computation, we
investigate the proximity effect and its influence on both the
superconducting order parameter and the local density of states in
the HM$\mid$S bilayer for a model system as described in the Fig.
\ref{fig:lattice}. The analysis is based on the case of a junction
configuration with the barrier aligned along the $y$ direction in
the $xy$ plane. The pairing amplitudes are expressed in terms of
the components along the $x$ and $y$ axes of the square lattice.
In particular, within such configuration, due to the symmetry
properties of the examined order parameters, we do expect that
Andreev bound states are effective only for the chiral $p$-wave
paired state (for the analysis on the lattice we do not
consider the case of $d_{xy}$ symmetry of the superconducting
order parameter).
The formalism to be used when calculating the order parameter profile and the
density of states is based on a lattice BdG-approach similar to that adopted
in Ref. \cite{cuoco_prb_08} but extended to the case of a spin active interface.
The total Hamiltonian $H$ of the system may be written as
\begin{align} H = H_F + H_S + H_T + H_I,
\end{align}
where $H_F$ and $H_S$ accounts for the ferromagnetic and superconducting layers, while
$H_T$ and $H_I$ describe the tunnelling processes and the scattering potential at the
interface region. We have
\begin{eqnarray}
H_l &= -\sum_{\langle \vi,\vj\rangle,\sigma} t_{l\sigma}
(c_{\vi\sigma}^\dag c_{\vj\sigma} + \text{h.c.}) + \sum_\vi U_A
n_{\vi\uparrow}n_{\vi\downarrow} \notag\\
&+ \sum_{\langle \vi,\vj\rangle}
V_A(n_{\vi\uparrow}n_{\vj\downarrow} +
n_{\vi\downarrow}n_{\vj\uparrow}) - \mu \sum_{\vi,\sigma}
n_{\vi\sigma} \notag\\
&- h_A\sum_{\vi\sigma} (n_{\vi\uparrow} - n_{\vi\downarrow}),\;
l=F,S, \label{Hext}
\end{eqnarray}
where $\langle \vi,\vj\rangle$ denotes nearest-neighbor sites,
$\{c_{\vi\sigma}^\dag,c_{\vi\sigma}\}$ are creation and
annihilation operators of an electron with spin $\sigma$ on site
$\vi=(i_x,i_y)$, while $n_{\vi\sigma}$ is the number operator. We
take the exchange field to be non-zero only on the ferromagnetic
side, where we let $h_F \to \mu$ in correspondence with our assumption of
a half-metallic limit. The hopping amplitudes are chosen such that
$t_{F\uparrow}= t_{F\downarrow}$=$t_{S\uparrow}=t_{S\downarrow}$=$t$.
Above, $\mu$ is the chemical potential, while $U_l$ and $V_l$ denote
on-site and nearest-neighbor interaction on side $l$.
\par The two layers communicate by means
of the tunneling Hamiltonian term, which reads
\begin{align}
H_T = -t_T\sum_{\langle \vi,\vj\rangle\sigma} (c_{\vi\sigma}^\dag
c_{\vj\sigma} + \text{h.c.}),
\end{align}
where the sites $\langle \vi,\vj\rangle$ are located at the
surface of the F and S layer. Finally, the scattering potential at
the interface is modelled through the term
\begin{align} H_I =
\sum_{\vi\alpha\beta} c_{\vi\alpha}^\dag [V_0\underline{1} +
\mathbf{V}_M\cdot\underline{\boldsymbol{\sigma}}]_{\alpha\beta}
c_{\vi\beta},
\end{align}
where $\{\alpha,\beta\}$ are spin-indices, while $\vi$ denotes a
lattice site on the surface of the F or S layer (for convenience
the interface has been placed at the site $x=0$ as indicated in
Fig. \ref{fig:lattice}). Here, $V_0$ is a spin-independent
scattering potential, roughly corresponding to the parameter $Z$
introduced previously, while $\mathbf{V}_M$ is a spin-dependent
scattering potential which gives rise to spin-mixing and spin-flip
processes. The Pauli-vector matrix is given as
$\underline{\boldsymbol{\sigma}} =
(\underline{\sigma_x},\underline{\sigma_y},\underline{\sigma_z})$
and $\underline{1}$ is the $2\times2$ identity matrix. To facilitate
comparison with the notation and parameters used when calculating the
conductance, we similarly define
\begin{align}
\mathbf{V}_M = -\rho
V_0(\cos\Psi\sin\phi,\sin\Psi\sin\phi,\cos\phi),
\end{align}
such that $\rho$ denotes the relative weight of the spin-independent and
spin-dependent potential while $\{\phi,\Psi\}$ provides the direction of
the magnetic moment at the interface.
\begin{figure}[b!]
\centering \resizebox{0.50\textwidth}{!}{
\includegraphics{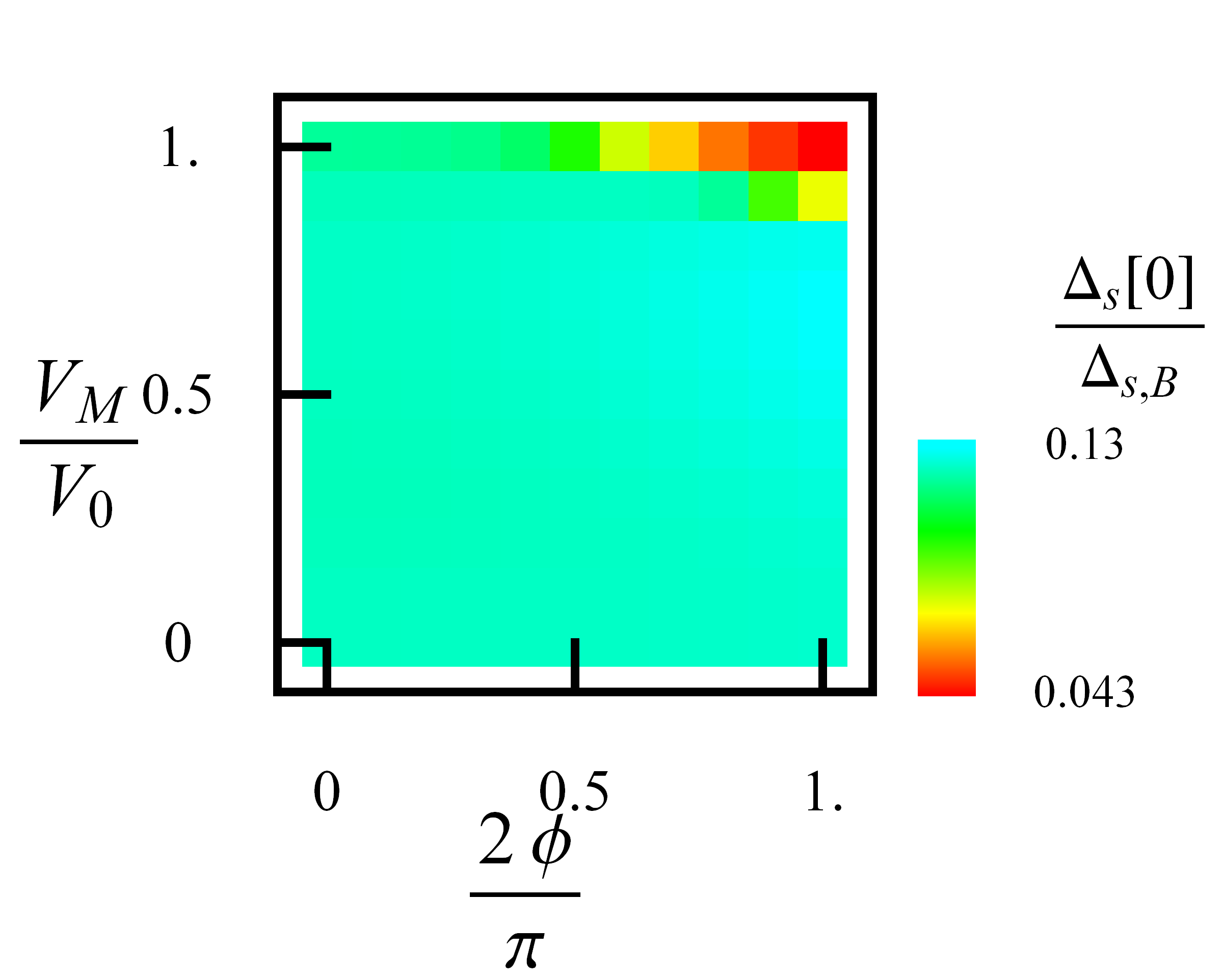}}
\caption{(Color online) Color map of the s-wave on-site pairing
amplitude at the interface $\Delta_S[0]$ with respect to the bulk
value $\Delta_{S,B}$ as a function of the misalignment angle
$\phi$ and the magnetic barrier strength $V_M$. The non magnetic
scattering potential has an amplitude $V_0=2 t$.
}\label{fig:figswT0}
\end{figure}
We analyze the effects of the spin-active barrier on the amplitude
and the phase of the superconducting order parameter by solving
the BdG equations on the lattice within the extended Hubbard model
introduced above. The calculation is performed for the case of a
planar bilayer junction of dimension 2$L_x \times$2$L_y$. We have
considered lattice sizes of $L_x$=$L_y$=\,40, 50, 60 (lattice
constant is the  the unit of length). The results obtained do not
show qualitative nor significant quantitative changes for these
values of $L_x,L_y$. The case stiudied here corresponds to
superconducting and magnetic coherence length of the order of
$\xi_S \simeq 7$ and $\xi_{HM} \simeq 1$. Hence, size effects can
be considered negligible for the systems we consider. Concerning
the ratio $\xi_S/\xi_{HM}$, other computations have been performed
at different values of the pairing strength $V$, thus varying
$\xi_S$. These cases do not show qualitative changes in the
results. In this framework, the modification of the coherence
length is limited by two conditions: i) the requirement of stable
superconducting solutions in the phase diagram for the order
parameters in the desired symmetry depends on the pair coupling,
and ii) the computational demand is related to the size of the
matrix Hamiltonian. Hereafter, the discussion will focus on the
case $L_x$=40. From Eq. \ref{Hext} and the BdG formalism already
described in Ref. \cite{cuoco_prb_08}, but extended to the case of
a spin active barrier, we compute the spatial variation of the
superconducting order parameter for different pairing symmetry and
as a function of the barrier parameters. Due to the presence of a
spin active barrier one has to introduce a four component
Bogoliubov basis on each atomic site to take into account both
particle-hole spin flip processes as well as the pairing channel
of particle-hole resonance. This introduces an extra factor in the
computational complexity. In particular, among the various results
obtained, the focus is on the modification, due to the split
exchange and the spin flip coupling, of the superconducting order
parameter evaluated at the HM|S interface versus $(\phi,V_M,V_0)$.
In doing that, we have to properly choose the interaction strength
both in the magnetic as well as in the superconducting subsystem
of the junction in order to get the desired microscopic quantum
states. To this end, the effective exchange amplitude in the
ferromagnetic region is taken as $h_F=4.0 t$ to have a
half-metallic behavior and a profile for the z-component of the
magnetization with zero spin minority carriers. Furthermore, to
get an $s$-wave, $d$-wave and a chiral $p$-wave symmetry within
the superconducting regioon, three different sets of attractive
pairing amplitudes and chemical potentials have to be
considered\cite{cuoco_prb_08}. For the onsite $s$-wave, we assume
a value of the chemical potential equal to $\mu=-0.2 t$, with
$U_F=0$, $V_F=0$ for the F side and $U_S=-1.5 t$, $V_S=0$ for the
S side. For the chiral $p$-wave, we choose the value $\mu=-1.6 t$,
with $U_F=0$, $V_F=0$ within the F side and $U_S=0$, $V_S=-2.5 t$
for the S side. For the $d$-wave, we choose $\mu=-0.2 t$, with
$U_F=0$, $V_F=0$ within the F side and $U_S=0$, $V_S=-2.5 t$ in
the S side.

Furthermore, the tunnelling matrix element is
kept fixed and chosen equal to $t_{T}=t$.

The interaction terms in $H_F$ and $H_S$ are decoupled by means of
a standard Hartree-Fock approximation such that the magnetic and
pairing channels originate from the on-site and the intersite
interactions, respectively:
\begin{eqnarray*}
U_F\, n_{\mathbf{i}\uparrow }n_{\mathbf{i}\downarrow } & \simeq &
U_F\, \left[ \langle n_{\mathbf{i}\downarrow }\rangle
n_{\mathbf{i}\uparrow }+\langle n_{\mathbf{i}\uparrow }\rangle
n_{\mathbf{i}\downarrow }-\langle n_{\mathbf{i}\uparrow }\rangle
\langle n_{\mathbf{i}\downarrow }\rangle
\right] \\
V_S\, n_{\mathbf{i}\uparrow }n_{\mathbf{j}\downarrow } & \simeq
&V_S\, \left[ \Delta _{\mathbf{ij}}\,c_{\mathbf{j}\downarrow
}^{\dagger }c_{\mathbf{i} \uparrow }^{\dagger
}+{\Delta}^*_{\mathbf{ij}}\, c_{\mathbf{i}\uparrow
}c_{\mathbf{j}\downarrow }-|\Delta
_{\mathbf{ij}}|^{2}\right] \\
U_S\, n_{\mathbf{i}\uparrow }n_{\mathbf{i}\downarrow } & \simeq
&U_S\, \left[ \Delta _{\mathbf{i}}\,c_{\mathbf{i}\downarrow
}^{\dagger }c_{\mathbf{i} \uparrow }^{\dagger
}+{\Delta}^*_{\mathbf{i}}\, c_{\mathbf{i}\uparrow
}c_{\mathbf{i}\downarrow }-|\Delta _{\mathbf{i}}|^{2}\right]\quad
.
\end{eqnarray*}
Here, we have introduced the on-site $\Delta _{\mathbf{i}}=\langle
c_{\mathbf{i}\uparrow }c_{\mathbf{i} \downarrow }\rangle $ and the
bond pairing amplitude on a bond $\Delta _{\mathbf{ij}}=\langle
c_{\mathbf{i}\uparrow }c_{\mathbf{j} \downarrow }\rangle $, with
the average $\langle K\rangle $ yielding the expectation value of
the operator $K$ over the ground state. Moreover, the on site
z-component $m_{z\mathbf{i}}= \frac{1}{2}(\langle
n_{\mathbf{i}\uparrow }\rangle -\langle n_{\mathbf{i} \downarrow
}\rangle )$ and the $(x,y)$-components of the magnetization
$m_{x\mathbf{i}}= \frac{1}{2}(\langle
c^{\dagger}_{\mathbf{i}\uparrow }c_{\mathbf{i} \downarrow } +h.c.
\rangle ), m_{y\mathbf{i}}= \frac{i}{2}(\langle
c^{\dagger}_{\mathbf{i}\uparrow }c_{\mathbf{i} \downarrow } -h.c.
\rangle )$ are iteratively determined up to the required accuracy
to get the spatial dependence of the vector spin polarization.
From the pairing amplitudes, it is possible to construct the
superconducting profiles for the different symmetries ($d$- and
$p$-wave) in the singlet (S) and triplet (T) channel in terms of
components of the z-projected axial spin operator. They are
defined as
\begin{eqnarray*}
\Delta _{d}(\mathbf{i}) & = & \left( \Delta
_{\mathbf{i},\mathbf{i+\hat{x }}}^{(S)}+\Delta
_{\mathbf{i},\mathbf{i-\hat{x}}}^{(S)}-\Delta _{\mathbf{i},
\mathbf{i+\hat{y}}}^{(S)}-\Delta _{\mathbf{i},\mathbf{i-\hat{y}}
}^{(S)}\right) /4 \\
\Delta _{p_{x}}(\mathbf{i}) & = & (\Delta _{\mathbf{i},
\mathbf{i+\hat{x}}}^{(T)}-\Delta
_{\mathbf{i},\mathbf{i-\hat{x}}}^{(T)})/2 \\
\Delta _{p_{y}}(\mathbf{i}) & = & (\Delta _{\mathbf{i},
\mathbf{i+\hat{y}}}^{(T)}-\Delta
_{\mathbf{i},\mathbf{i-\hat{y}}}^{(T)})/2
\end{eqnarray*}
for $d_{x^{2}-y^{2}}$, $p_{x}$ and $p_{y}$-wave, respectively. Here one has
to introduce the singlet and triplet pairing amplitudes on a bond, given by
\begin{eqnarray*}
\Delta _{\mathbf{ij}}^{S} & = & \left( \Delta_{\mathbf{ij}}+\Delta
_{\mathbf{ji}}\right) /2 \\
\Delta _{\mathbf{ij}}^{T} & = & \left( \Delta
_{\mathbf{ij}}-\Delta _{\mathbf{ji}}\right) /2 \quad .
\end{eqnarray*}
We adopt open (periodic) boundary conditions for the direction $x$
perpendicular ($y$ parallel) to the interface, taking the Fourier
transform due to the translational invariance along the $y$
direction of the relevant physical quantities.

\begin{figure}[b!]
\centering \resizebox{0.50\textwidth}{!}{
\includegraphics{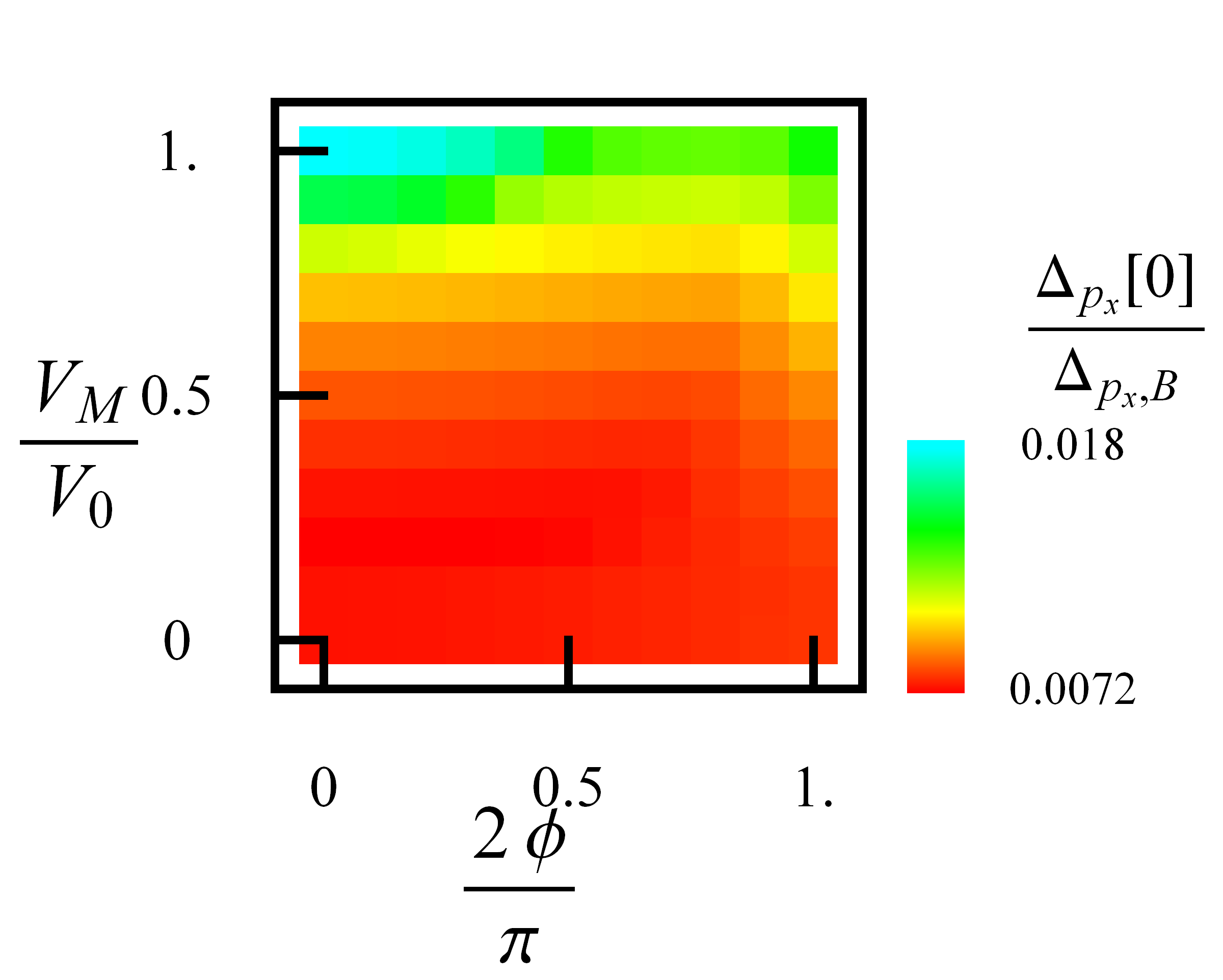}}
\caption{(Color online) Color map of the real part of the $p_x$
component of the pairing amplitude evaluated at the interface
$\Delta_{p_x}[0]$ with respect to the bulk value $\Delta_{p_x,B}$
as a function of the scaled misalignment angle $2\phi/\pi$ and the
magnetic barrier strength $V_M/V_0$. The non magnetic scattering
potential has a given amplitude of $V_0=2 t$.
}\label{fig:figpxwT0}
\end{figure}
%%%%%%%%%
\begin{figure}[b!]
\centering \resizebox{0.50\textwidth}{!}{
\includegraphics{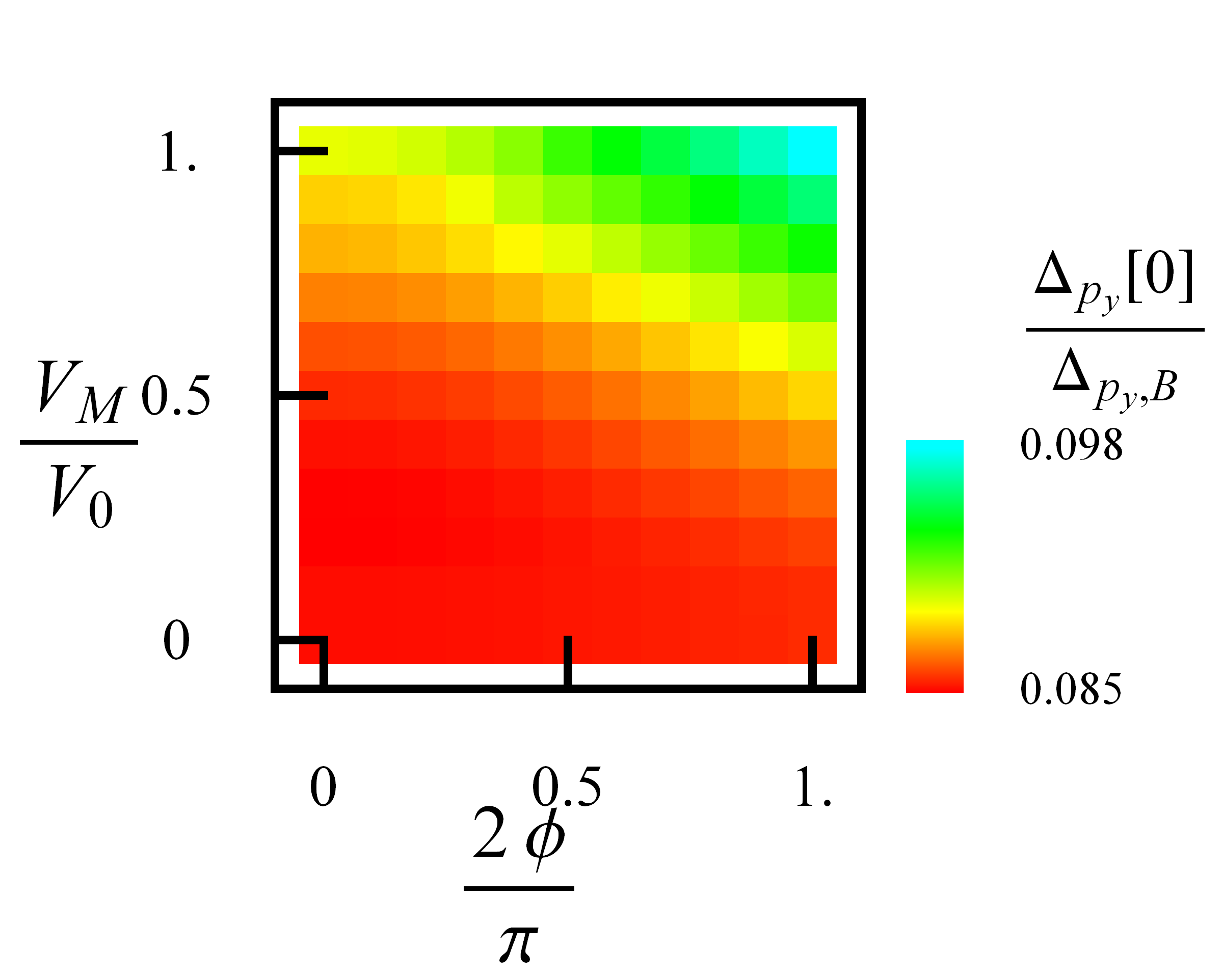}}
\caption{((Color online) Color map of the imaginary part of the
$p_y$ component for the pairing amplitude evaluated at the
interface $\Delta_{p_y}[0]$ with respect to the bulk value
$\Delta_{p_y,B}$ as a function of the scaled misalignment angle
$2\phi/\pi$ and the ratio of the magnetic barrier strength with
respect to the nonmagnetic one, $V_M/V_0$. The non magnetic
scattering potential has an amplitude $V_0=2 t$.
}\label{fig:figpywT0}
\end{figure}
%%%%%%%%%%
\begin{figure}[b!]
\centering \resizebox{0.50\textwidth}{!}{
\includegraphics{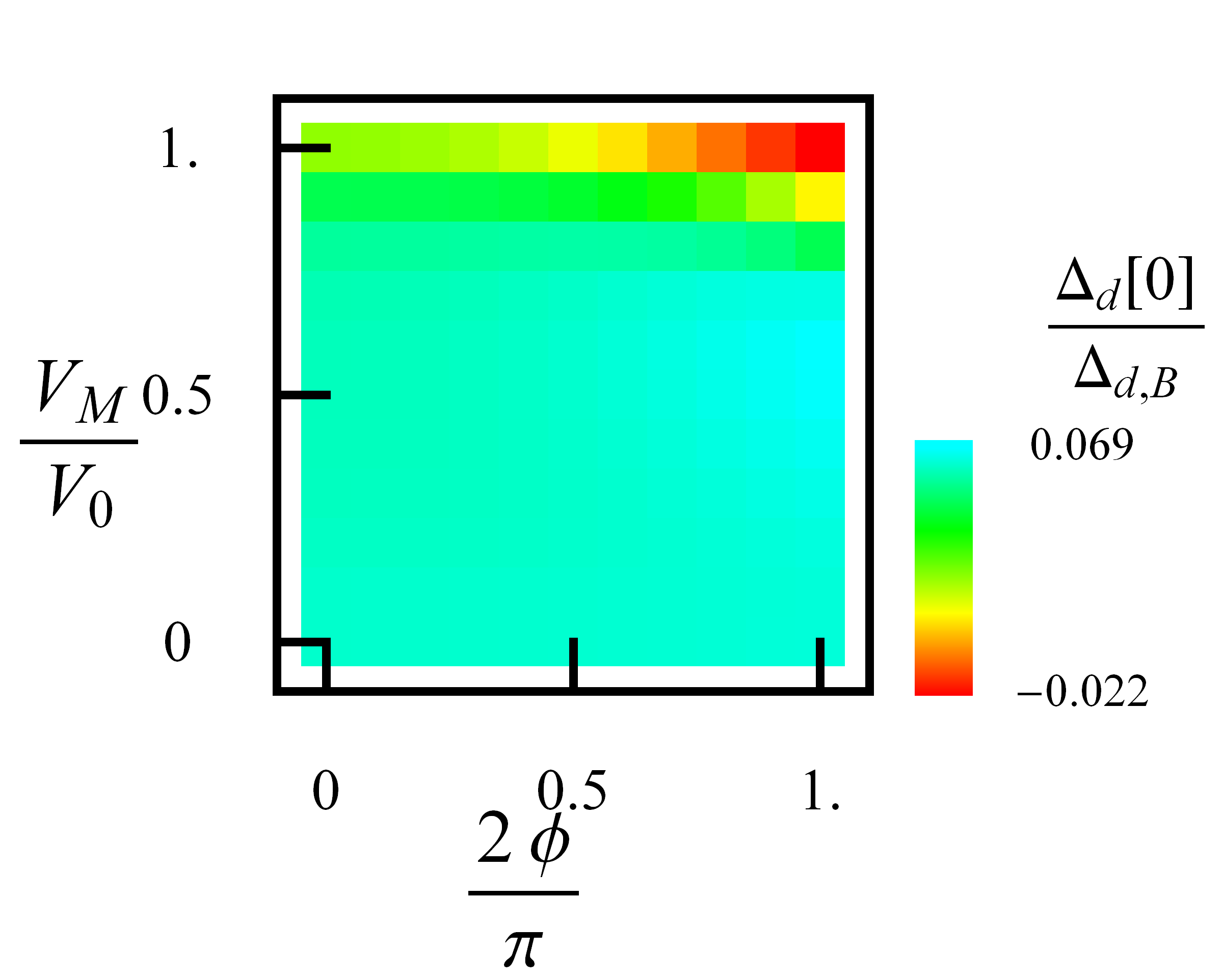}}
\caption{(Color online) Color map the $d_{x^2-y^2}$ pairing
amplitude at the interface $\Delta_{d}[0]$ with respect to the
bulk value $\Delta_{d,B}$ as a function of the scaled misalignment
angle $2\phi/\pi$ and the ratio of the magnetic barrier strength
with respect to the non-magnetic one, $V_M/V_0$. The non magnetic
scattering potential has an amplitude $V_0=2 t$.
}\label{fig:figdwT0}
\end{figure}
%%%%%%%%%%
In order to understand the role of the spin active barrier in
tuning amplitude and phase of the different symmetry order
parameters we have investigated their evolution in terms of the
barrier parameters at zero temperature. The angle $\phi$ tunes the
direction of the barrier spin moment from parallel to the
$z$-quantization axis at $\phi=0$ to being parallel to the $x$-
direction for $\phi=\pi/2$. Different values of $\Psi$ do not
change the results due to the $x-y$ symmetry in the spin space.
Thus, it suffices to consider the case $\Psi=0$.
Furthermore, we have chosen one representative case for the regime
of nonmagnetic barrier strength that corresponds to a situation of
reduced electron density at the interface compared to the
electron distribution in the HM and S sides. In this respect, the
way of introducing a scattering potential at the interface may
lead to extra effects if compared to the delta potential
considered in the BTK formalism, since the amplitude of the non
magnetic potential determines the average electron occupation at
the barrier site as well as in its proximity.

{\it s-wave pairing} As one can see in Fig. \ref{fig:figswT0}, the
value of the s-wave order parameter does not vary significantly
in the full range of values for $V_M$ and $\phi$. Thus, the amplitude reduction
with respect to the bulk value is basically controlled by the
presence of the half-metallic ferromagnet. Only by approaching the
regime of $V_M \simeq V_0$ the effects of the spin active barrier
become more relevant leading to a strong suppression of the
pairing amplitude. In this case it is possible to distinguish two
different behaviors corresponding to the spin-mixing or spin-flip
barrier regime. Spin-mixing effects (i.e. $\phi=0$) are not much
relevant for the s-wave proximity effect as the pair amplitude
exhibits only a slight reduction as one tunes the spin-dependent
scattering from the regime $\rho=0$ to $\rho\simeq 1$. This can be
understood because the change of $V_M$ tends to reinforce the
magnetization even at the barrier site. There, the proximity between
the half-metal ferromagnet and the superconductor leads to a
matching of the magnetization from full- to zero- spin
polarization in moving through the interface. Otherwise, spin-flip
mechanisms lead to a larger reduction of the pair amplitude when
approaching the limit $\rho\rightarrow 1$. In this case, the
increase of the transverse magnetization (parallel to $x$ in the
spin space) at the barrier site leads to extra scattering for the
singlet pairs that in turn sums up to the pair breaking effect due
to the presence of the half-metal ferromagnet in a way to get
about a 70$\%$ reduction.

{\it Chiral p-wave pairing} Consider next the case of
chiral $p$-wave pairing (i.e. $\Delta \sim $ $p_x+i\,p_y$) on
the superconducting side of the junction. The chiral state exhibits
time reversal symmetry breaking for a spin triplet configuration. It is
well known that the interface properties can be quite unusual even at
the boundary with the vacuum due the possibility of emergent exotic
edge states. Here, we analyze the consequences of spin-mixing and spin-flip
barrier processes on the two $p$-wave components separately. Though
the gap amplitude is isotropic in $k$-space, as for the
$s$-wave symmetry, the pairing amplitudes along $x$ and $y$ evaluated
at the interface exhibit a completely different behavior. There are many
distinguishing features that can be extracted from inspecting Figs.
\ref{fig:figpxwT0} and \ref{fig:figpywT0}. Regarding effects induced by
spin-mixing and spin-flip processes, note that i) spin-mixing and
spin-flip reduce the pair breaking effects due to the proximity
with the half-metal, ii) the reduction is not equivalent for the
$x$ and $y$ components, iii) the pair breaking effects are more
pronounced for the $p_x$ than the $p_y$ component. Indeed, for the
case i) one notices in Figs. \ref{fig:figpxwT0} and
\ref{fig:figpywT0} that in the regime $V_M/V_0 \sim 1$, the pair
amplitude tends to grow at any given angle with a slope that is
more pronounced for the $p_x$ component as compared to the $p_y$
one. Concerning the point ii), the maximum of the pairing
amplitude occurs at the phase diagram positions individuated by
$(\phi,V_M/V_0)=(0,1)$ and $(\phi,V_M/V_0)=(\pi/2,1)$ for the
$p_x$ and the $p_y$ amplitude, respectively. We argue that the
presence of Andreev bound states in the spectrum, due to the
change of sign of the $p_x$ component in the direction
perpendicular to the interface, leads to a more significant
barrier influence of the correspondent pairing component with
respect to the $p_y$ one.
\begin{figure}[b!]
%\centering \resizebox{0.45\textwidth}{!}{
%\includegraphics{dospwUint05.png}}
\centering \resizebox{0.5\textwidth}{!}{
\includegraphics{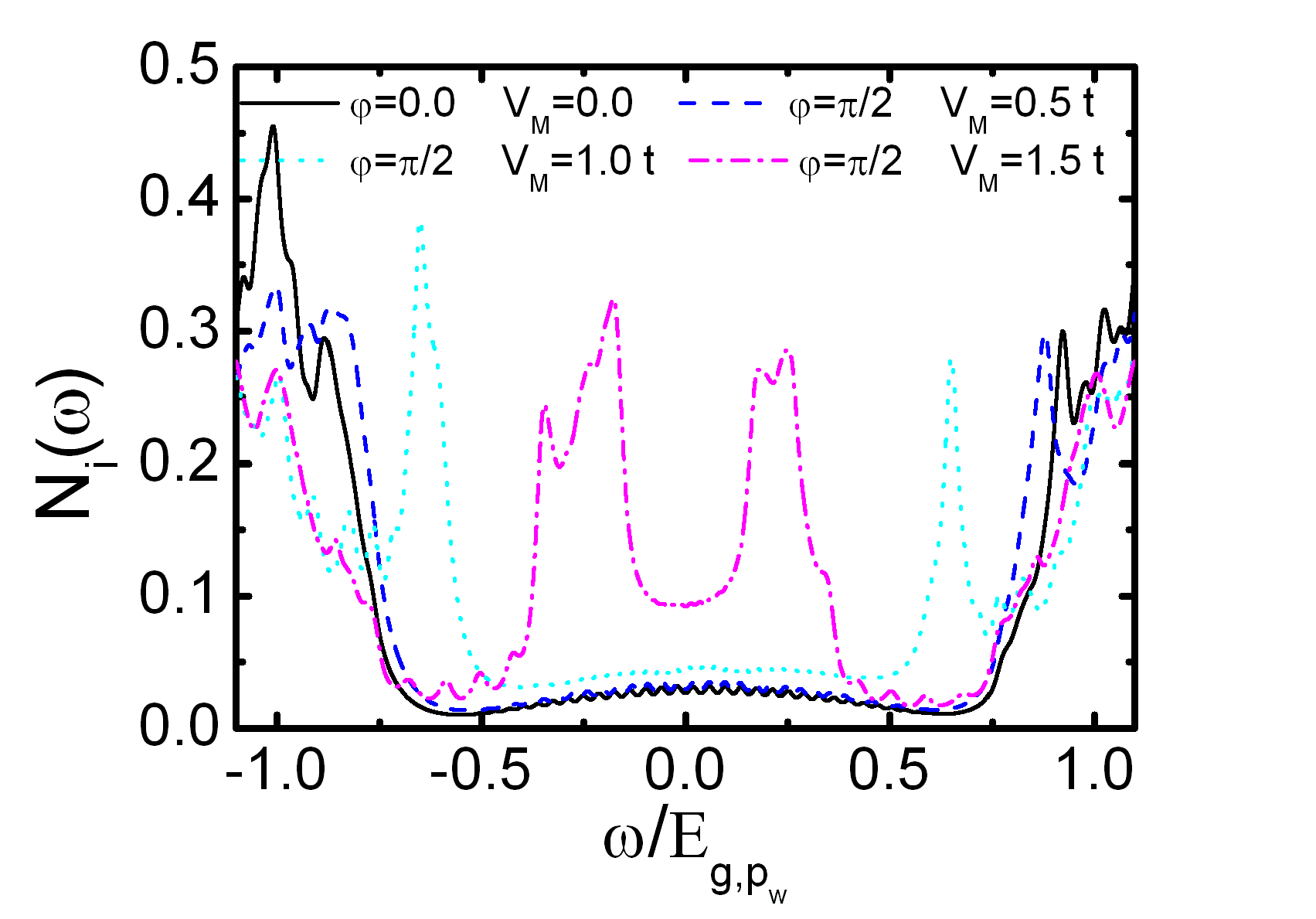}}
\caption{(Color online) Density of states for the chiral $p$-wave
evaluated at one representative site position ($i=2$) within the
range of one superconducting coherence length from the barrier
($i=0$). The non magnetic potential has a value of $V_0=0.5$t.
}\label{fig:dospwUint0.5}
\end{figure}

{\it d-wave pairing} In Fig. \ref{fig:figdwT0}, we show the
evolution of the interface d-wave pairing amplitude in terms of
the scaled angle and scattering barrier parameters. Recall that
for the chosen junction geometry there are no Andreev bound states
in proximity to the half-metal. Hence, even though the
pairing is anisotropic, the absence of Andreev states leads us
to expect a behavior similar to the $s$-wave case. Indeed, the
profile of the $d$-wave pairing amplitude at the interface is
similar to that of the isotropic $s$-wave with an increase of the
pair breaking effects in the regime of large spin-mixing and
spin-flip. However, it is worth pointing out that close to the
regime of maximal spin-flip scattering ($\phi=\pi/2$), where the
orientation of the magnetic moment at the interface is perpendicular
to the easy axis of pair formation, the combination of the spin-flip
and the presence of the half-metal leads to a global phase change
in the d-wave amplitude. This is reminiscent of the oscillating
behaviour one would expect in a conventional ferromagnet/superconductor
junction. Here, it is the anisotropy of the pairing that favors the formation
of small oscillations in proximity of the barrier.

{\it{Interface density of states (DOS)}} In the previous section,
we saw how a finite subgap conductance was obtained even in the
strong tunnelling limit when the superconducting pairing was
unconventional and the interface allowed for spin-flip processes.
We speculated that the physical explanation behind this phenomenon
was the generation of surface bound-states, appearing due to an
interplay between the spin-active properties of the interface and
the internal phase of the superconducting order parameter. In the
absence of a spin-active interface, we found that subgap
tunnelling vanished completely. We now investigate whether such
surface bound-states are truly present or not within the lattice
BdG-model, by focusing on the p-wave chiral type of pairing. This
is expected to yield Andreev bound states for the geometry given
in Fig.\ref{fig:lattice}.

In Fig.\ref{fig:dospwUint0.5}, we show the behavior of the
total DOS
\begin{eqnarray*}
N_{i}(\omega)= N_{i\uparrow}(\omega)+N_{i\downarrow}(\omega)
\end{eqnarray*}
evaluated at one representative site position within a range of
one superconducting coherence length from the barrier. There is a
substantial difference in the results for the total DOS obtained
when comparing the case of a complete
non-magnetic barrier ($\phi=0$ and $V_M=0$) with the case of a
pure spin-flip barrier scattering potential ($\phi=\pi/2$ and a
varying $V_M$). The energy has been rescaled with respect to the
gap amplitude $E_{g,p_w}$ determined inside the superconductor
within the same formalism, that is at a position where the DOS
exhibits a full gap as expected for the chiral p-wave symmetry and
the order parameter is uniform and unaffected by the
interfaces. Note that for non-zero $\phi$ the change in
the amplitude of $V_M$ leads to extra midgap edge states. The
energy of these states depend strongly on the
magnetic barrier potential. Indeed, they appear at the edge of the
gap for small values of $V_M$ and as the amplitude of the magnetic
moment at the barrier increases by tuning $V_M$, they
shift towards low energies. The asymmetry of the resonant spectra is
related to the presence of a small, but finite spin polarization
proximate to the barrier on the superconducting side.

\section{Summary}\label{sec:summary}

In summary, by means of continuum and lattice BdG formalisms we
have computed the conductance spectra, superconducting order parameter,
and the density of states of a bilayer system made of a half-metal
and an unconventional superconductor which are brought into contact through
a spin active tunneling. These quantities  have been computed for different
forms of the pairing amplitude both for spin singlet $s$- and $d$-wave
as well as spin triplet chiral $p$-wave symmetry. We have shown how the
spin-flip and spin-mixing scattering processes at the interface influence the structure
of the superconducting order parameter in the case of different pairing symmetries,
namely $s$-wave, chiral $p$-wave, and $d$-wave pairing. These scattering
prcoesses lead to different charge transport features such as subgap conductance
and midgap Andreev assisted tunnelling due to resonant states, depending
on which pairing symmetries are considered. The novel subgap features we find
are present only for a non-zero misalignment between the
half-metallic magnetization and the magnetic moment of the barrier.
The energy position of the Andreev assisted charge processes turns out to be
sensitive to changes in the ratio between the magnetic and
non-magnetic scattering potential at the interface.
We have also studied the pairing amplitude in
a lattice model and computed its dependence at the
interface on the barrier scattering strength, as well as its
dependence on the misalignment angle between the half-metal magnetization and the
barrier spin moment.
%Different pairing symmetries (we have considered s-wave, chiral
%p-wave, and d-wave symmetries) respond qualitatively differently
%to spin-mixing and spin-flip scattering at the
%interface.
For the case of a spin triplet chiral $p$-wave, midgap
states at non zero misalignment angle between the half-metal magnetization
the barrier spin moment have been  found  by computing the density of
states of the superconductor close to the interface. A hallmark of such
edge states is the strong dependence of the ratio between the magnetic and
non-magnetic scattering potential in the subgap DOS. Finally, we point out
that the present study reveals highly non-trivial features of the proximity
effect between a half-metal and an unconventional superconductor in the
presence of a spin active interface even without invoking the occurrence
of exotic mixed parity pair components. The present analysis is the
starting point for further investigation of the role played by
induced or subdominant pairing amplitudes in heterostructures
based on half-metal and unconventional superconductors.

\acknowledgments T. Yokoyama is acknowledged for useful discussions. J.L. and
A.S. were supported by the Norwegian Research Council Grant No. 167498/V30
(STORFORSK).

\end{document}